\documentclass[italian,english]{article}
\usepackage[latin9]{inputenc}
\usepackage{color}
\usepackage{amssymb}
\usepackage{esint}

\makeatletter
\newcommand{\lyxaddress}[1]{
\par {\raggedright #1
\vspace{1.4em}
\noindent\par}
}

\makeatother

\usepackage{babel}

\begin{document}

\title{\textbf{Irreversible gravitational collapse: black stars or black
holes? }}

\author{\textbf{\Large Christian Corda$^{0,4}$ and Herman J. Mosquera Cuesta$^{1,2,3,4}$ }}

\maketitle

\lyxaddress{\begin{center}
\textbf{$^{0}$}Institute for Basic Research, P. O. Box 1577, Palm
Harbor, FL 34682, USA%
\footnote{\begin{quotation}
CC is partially supported by a Research Grant of The R. M. Santilli
Foundation Number RMS-TH-5735A2310
\end{quotation}
}
\par\end{center}}

\lyxaddress{\begin{center}
\textbf{$^{1}$}Departmento de Fìsica Universidade Estadual Vale
do Acaraù, Avenida da Universidade 850, Campus da Betània, CEP 62.040-370,
Sobral, Cearà, Brazil%
\footnote{HJMC is fellow of the Cearà State Foundation for the Development of
Science and Technology (FUNCAP), Fortaleza, CE, Brazil%
}
\par\end{center}}

\lyxaddress{\begin{center}
$^{2}$Instituto de Cosmologia, Relatividade e Astrofìsica (ICRA-BR),
Centro Brasilero de Pesquisas Fìsicas, Rua Dr. Xavier Sigaud 150,
CEP 22290 -180 Urca Rio de Janeiro, RJ, Brazil
\par\end{center}}

\begin{center}
$^{3}$International Center for Relativistic Astrophysics Network
(ICRANet), International Coordinating Center, Piazzale della Repubblica
10, 65122, Pescara (PE), Italy
\par\end{center}

\lyxaddress{\begin{center}
\textbf{$^{4}$}International Institute for Theoretical Physics and
Mathematics Einstein-Galilei, Via Santa Gonda, 14 - 59100 Prato, Italy
\par\end{center}}

\begin{center}
\textit{E-mail addresses:} \textcolor{blue}{cordac.galilei@gmail.com,
herman@icra.it}
\par\end{center}
\begin{abstract}
{\normalsize It is well known that the concept of black hole has been
considered very fascinating by scientists even before the introduction
of Einstein's general relativity. They should be the final result
of an irreversible gravitational collapse of very massive bodies. }{\normalsize \par}

{\normalsize However, an unsolved problem concerning such objects
is the presence of a space-time singularity in their core. Such a
problem was present starting by the first historical papers concerning
black holes. It is a common opinion that this problem could be solved
when a correct quantum gravity theory will be, finally, constructed. }{\normalsize \par}

{\normalsize In this work we review a way to remove black hole singularities
at a classical level i.e. without arguments of quantum gravity. By
using a particular non-linear electrodynamics Lagrangian, an exact
solution of Einstein field equations is shown. The solution prevents
the collapsing object to reach the gravitational radius, thus the
final result becomes a black star, i.e. an astrophysical object where
both of singularities and event horizons are removed. Such solution
is not only a mathematical artifice. In fact, this kind of Lagrangian
has been recently used in various analysis in astrophysics, like surface
of neutron stars and pulsars. The authors also recently adapted the
analysis on a cosmological context by showing that the big-bang singularity
can be removed too. }{\normalsize \par}
\end{abstract}

\begin{quotation}
This paper is dedicated to the Memory of Professor Darryl Jay Leiter,
February 25, 1937 - March 4, 2011. Professor Leiter gave a fundamental
contribution in evolving an alternate explanation of black holes,
the theory of MECOs or magnetic eternally collapsing compact objects. 
\end{quotation}

\section{{\large Introduction}}

The concept of black-hole (BH) has been considered very fascinating
by scientists even before the introduction of general relativity (see
\cite{key-1} for an historical review). A BH is a region of space
from which nothing, not even light, can escape out to infinity. It is the result of
the deformation of spacetime caused by a very compact mass. Around
a BH there is an undetectable surface which marks the point of no
return. This surface is called an event horizon. It is called \char`\"{}black\char`\"{}
because it absorbs all the light that hits on it, reflecting nothing,
just like a perfect black body in thermodynamics \cite{key-2}. However,
an unsolved problem concerning such objects is the presence of a space-time
singularity in their core. Such a problem was present starting by
the first historical papers concerning BHs \cite{key-3,key-4,key-5}.
It is a common opinion that this problem could be solved when a correct
quantum gravity theory will be, finally, obtained, see \cite{key-6}
for recent developments in this direction.

On the other hand, fundamental issues which dominate the question
about the existence or non-existence of BH horizons and singularities
and some ways to avoid the development of BH singularities within
the classical theory, which does not require the need for a quantum
gravity theory, have been discussed by various authors in the literature,
see references from \cite{key-7} to \cite{key-16}. In fact, by considering
the exotic nature of BHs, it may be natural to question if such bizarre
objects could indeed exist in nature or rather to suggest that they are merely pathological
solutions to Einstein's equations. Einstein himself thought that BHs
would not form, because he held that the angular momentum of collapsing
particles would stabilize their motion at some radius \cite{key-17}.

Let us recall some historical notes. In 1915, A. Einstein developed
his theory of general relativity \cite{key-18}. A few months later,
K. Schwarzschild gave the solution for the gravitational field of
a point mass and a spherical mass \cite{key-3}. A few months after
Schwarzschild, J. Droste, a student of H. Lorentz, independently gave
an apparently different solution for the point mass and wrote more
extensively about its properties \cite{key-19}. In such a work Droste
also claimed that his solution was physically equivalent to the one
by Schwarzschild. In the same year, 1917, H. Weyl re-obtained the
same solution by Droste \cite{key-20}. This solution had a peculiar
behaviour at what is now called the Schwarzschild radius, where it
became singular, meaning that some of the terms in the Einstein equations
became infinite. The nature of this surface was not quite understood
at the time, but Hilbert \cite{key-21} claimed that the form by Droste
and Weyl was preferable to that in \cite{key-3} and ever since then
the phrase \textquotedblleft{}\emph{Schwarzschild solution}\textquotedblright{}
has been taken to mean the line-element which was found in \cite{key-19,key-20}
rather than the original solution in \cite{key-3}. 

For the sake of completeness we recall that, based on new translations
of Schwarzschild's original work, there are researchers who invoke
the non existence of BHs by claiming that the Schwarzschild's original
work \cite{key-3} gives a solution which is physically different
from the one derived by Droste \cite{key-19} and Weyl \cite{key-20}.
The new translations of Schwarzschild's original work can be found
in ref. \cite{key-22,key-23}. These works commented on Schwarzschild's
original paper \cite{key-3}. In particular Abrams \cite{key-22}
claimed that the line-element (we use natural units in all this paper)

\begin{equation}
ds^{2}=(1-\frac{r_{g}}{r})dt^{2}-r^{2}(\sin^{2}\theta d\varphi^{2}+d\theta^{2})-\frac{dr^{2}}{1-\frac{r_{g}}{r}}\label{eq: Hilbert}\end{equation}
 i.e. the famous and fundamental solution to the Einstein field equations
in a vacuum, gives rise to a space-time that is neither equivalent to
Schwarzschild's original solution in \cite{key-3}. In a following
work \cite{key-24} Abrams further claimed that {}``\emph{Black Holes
are The Legacy of Hilbert's Error}'' as Hilbert's derivation used
a wrong variable. Thus, Hilbert's assertion that the form of (\ref{eq: Hilbert})
was preferable to the original one in \cite{key-3} should be misleading.
Based on this, there are plenty of authors who agree with Abrams by claiming
that the work of Hilbert was wrong and Hilbert's mistake spawned the
BHs and the community of theoretical physicists continues to elaborate
on this falsehood, with a hostile shouting down of any and all voices
challenging them, see for example references \cite{key-23,key-25}.
In any case, this issue has been ultimately clarified in \cite{key-26}
where it has been shown that \emph{{}``the original Schwarzschild
solution''} \cite{key-3} results physically equivalent to the solution
(\ref{eq: Hilbert}) enabled like the correct one by Hilbert in \cite{key-21},
i.e. the solution that is universally known like the \char`\"{}Schwarzschild
solution\char`\"{} \cite{key-1}. The authors who claim that the original
Schwarzschild solution leaves no room for the science fiction of the
BHs (see references from \cite{key-22} to \cite{key-25}) give the
wrong answer \cite{key-26}. The misunderstanding is due to an erroneous
interpretation of the different coordinates \cite{key-26}. In fact,
arches of circumference appear to follow the law\emph{ }$dl=rd\varphi$,
if the origin of the coordinate system is a non-dimensional material
point in the core of the BH, while they do not appear to follow such
a law, but to be deformed by the presence of the mass of the central
body $M$ if the origin of the coordinate system is the surface of
the Schwarzschild sphere, see \cite{key-26} for details.

After this clarification, let us return on historical notes. In 1924,
A. Eddington showed that the singularity disappeared after a change
of coordinates (Eddington coordinates \cite{key-27}), although it
took until 1933 for G. Lemaître to realize, in a series of lectures
together with Einstein, that this meant the singularity at the Schwarzschild
radius was an unphysical coordinate singularity \cite{key-28}.

In 1931, S. Chandrasekhar calculated that a non-rotating body of electron-degenerate
matter above 1.44 solar masses (the Chandrasekhar limit) would collapse
\cite{key-5}. His arguments were opposed by many of his contemporaries
like Eddington, Lev Landau and the same Einstein. In fact, a white
dwarf slightly more massive than the Chandrasekhar limit will collapse
into a neutron star which is itself stable because of the Pauli exclusion
principle \cite{key-1}. But in 1939, \foreignlanguage{italian}{J.
R. Oppenheimer and G. M. Volkoff} predicted that neutron stars above
approximately 1.5 - 3 solar masses (the famous Oppenheimer\textendash{}Volkoff
limit) would collapse into BHs for the reasons presented by Chandrasekhar,
and concluded that no law of physics was likely to intervene and stop
at least some stars from collapsing to BHs \cite{key-29}. Oppenheimer
and \foreignlanguage{italian}{Volkoff} interpreted the singularity
at the boundary of the Schwarzschild radius as indicating that this
was the boundary of a bubble in which time stopped. This is a valid
point of view for external observers, but not for free-falling observers.
Because of this property, the collapsed stars were called \char`\"{}\emph{frozen
stars}\char`\"{} \cite{key-30} because an outside observer would
see the surface of the star frozen in time at the instant where its
collapse takes it inside the Schwarzschild radius. This is a known
property of modern BHs, but it must be emphasized that the light from
the surface of the frozen star becomes redshifted very fast, turning
the BH black very quickly. Originally, many physicists did not accept
the idea of time standing still at the Schwarzschild radius, and there
was little interest in the subject for lots of time. But in 1958,
D. Finkelstein, by re-analysing Eddington coordinates, identified
the Schwarzschild surface $r=2M$ (in \emph{natural units}, i.e. $G=1$,
$c=1$ and $\hbar=1$, i.e where $r$ is the radius of the surface
and $M$ is the mass of the BH) as an \emph{event horizon}, \char`\"{}\emph{a
perfect unidirectional membrane: causal influences can cross it in
only one direction}\char`\"{} \cite{key-31}. This extended Oppenheimer's
results in order to include the point of view of free-falling observers.
Finkelstein's solution extended the Schwarzschild solution for the
future of observers falling into the BH. Another complete extension
was found by M. Kruskal in 1960 \cite{key-32}. 

These results generated a new interest on general relativity, which,
together with BHs, became mainstream subjects of research within the
Scientific Community. This process was endorsed by the discovery of
pulsars in 1968 \cite{key-33} which resulted to be rapidly rotating
neutron stars. Until that time, neutron stars, like BHs, were regarded
as just theoretical curiosities; but the discovery of pulsars showed
their physical relevance and spurred a further interest in all types
of compact objects that might be formed by gravitational collapse.

In this period more general BH solutions were found. In 1963, R. Kerr
found the exact solution for a rotating BH \cite{key-34}. Two years
later E. T. Newman and A. Janis found the asymmetric solution for
a BH which is both rotating and electrically charged \cite{key-35}.
Through the works by W. Israel, B. Carter and D. C. Robinson the no-hair
theorem emerged \cite{key-1}, stating that a stationary BH solution
is completely described by the three parameters of the Kerr\textendash{}Newman
metric; mass, angular momentum, and electric charge \cite{key-1}.

For a long time, it was suspected that the strange features of the
BH solutions were pathological artefacts from the symmetry conditions
imposed, and that the singularities would not appear in generic situations.
This view was held in particular by Belinsky, Khalatnikov, and Lifshitz,
who tried to prove that no singularities appear in generic solutions
\cite{key-1}. However, in the late sixties R. Penrose and S. Hawking
used global techniques to prove that singularities are generic \cite{key-1}.

The term \char`\"{}\emph{black hole}\char`\"{} was first publicly
used by J. A. Wheeler during a lecture in 1967 \cite{key-36} but
the first appearing of the term, in 1964, is due to A. Ewing in a
letter to the American Association for the Advancement of Science
\cite{key-37}, verbatim: {}``\emph{According to Einstein\textquoteright{}s
general theory of relativity, as mass is added to a degenerate star
a sudden collapse will take place and the intense gravitational field
of the star will close in on itself. Such a star then forms a \textquoteleft{}black
hole\textquoteright{} in the universe}.'' 

In any case, after Wheeler's use of the term, it was quickly adopted
in general use. 

Today, the majority of researchers in the field is persuaded that
there is no obstacle to forming an event horizon. On the other hand,
there are other researchers who demonstrated that various physical
mechanisms can, in principle, remove both of event horizon and singularities
during the gravitational collapse \cite{key-7} - \cite{key-16}.
In particular, in \cite{key-9} an exact solution of Einstein field
equations which removes both the event horizon and singularity has
been found by constructing the right-hand side of the field equations,
i.e. the stress-energy tensor, through a non-linear electrodynamics
Lagrangian which was previously used in studying super-strongly magnetized compact
objects, in particular neutron stars and pulsars \cite{key-38,key-39}.
In the next Section we will discuss this important issue.

\section{{\large Non-singular gravitational collapse}}

In Einstein's General Theory of Relativity the Einstein equation relates
the curvature tensor of spacetime on the left hand side to the energy-momentum
tensor in spacetime on the right hand side \cite{key-1,key-40}. Within
the context of the Einstein equation the \emph{strong principle of
equivalence} (SPOE) requires that special relativity must hold locally
for all of the laws of physics in all of spacetime as seen by time-like
observers (\cite{key-40} and Section 2.1 of \cite{key-41}). Hence,
in the context of the SPOE this implies that the frames of reference
of co-moving observers within a gravitationally collapsing object
are required to always be able to be connected to the frame of reference
of stationary observers by special relativistic transformations with
physical velocities which are less than the speed of light in a vacuum \cite{key-40}.
Recently plausible arguments have been made which support the idea
that physically acceptable solutions to the Einstein equation will
only be those which preserve the SPOE as \emph{a law of nature} in
the universe \cite{key-7,key-8,key-16,key-40}. The observable consequence
of preserving the SPOE as a \emph{a law of nature} would be that compact
objects which emerge from the process of gravitation collapse could
not have event horizons (EHs) because their existence would prevent
co-moving observers within a gravitationally collapsing object from
being able to be connected to the frame of reference of stationary
observers by special relativistic transformations with physical velocities
which are less than the speed of light \cite{key-40}. Hence, as a
result of the SPOE, objects having EHs with non-zero mass would be
physically prohibited \cite{key-7,key-8,key-16,key-40}. In particular,
the preservation of the SPOE in the Einstein equation would put an
overall constraint on the nature of the non-gravitational physical
elements which go into the energy-momentum tensor on the right hand
side of the Einstein equation. However this constraint would not uniquely
determine the specific form of the non-gravitational dynamics of the
energy-momentum tensor \cite{key-7,key-8,key-16,key-40}. For this
reason many different theories can be constructed (e.g. \emph{eternally
collapsing objects} (ECO), \emph{magnetospheric eternally collapsing
objects} (MECO), \emph{nonlinear electrodynamics} (NLED) extremely compact \emph{objects},
which preserve the SPOE and hence can generate highly redshifted compact
objects without EHs \cite{key-7,key-8,key-9,key-16,key-40}. Since
each of these different SPOE preserving theories have unique observational
predictions associated with the interaction of their non-gravitational
components with the environment of their highly redshifted compact
objects without EHs, the specific one chosen by Nature can only be
determined by astrophysical observations which test these predictions
\cite{key-7,key-8,key-16,key-40}. In the following, we will review the 
NLED model in \cite{key-9}.

NLED Lagrangian has been used in various analysis in astrophysics,
like the surface of neutron stars \cite{key-38} and pulsars \cite{key-39},
and also on cosmological context to remove the big-bang singularity
\cite{key-42,key-43}. 

The effects arising from a NLED become quite important in super-strongly
magnetized compact objects, such as pulsars and particular neutron
stars \cite{key-38,key-39}. Some examples include the so-called magnetars
and strange quark magnetars. In particular, NLED modifies in a fundamental
basis the concept of gravitational redshift as compared to the well
established method introduced by standard treatments \cite{key-38}.
The analyses proved that, unlike using standard linear electrodynamics,
where the gravitational redshift is independent of any background
magnetic field, when a NLED is incorporated into the photon dynamics,
an effective gravitational redshift appears, which happens to depend
decidedly on the magnetic field pervading the pulsar. An analogous
result has also been obtained for magnetars and strange quark magnetars
\cite{key-39}. The resulting gravitational redshift tends to infinity
as the magnetic field grows larger \cite{key-22,key-23}, as opposed
to the predictions of standard analyses which involve linear electrodynamics.
What  is important here is that the gravitational redshift of neutron
stars is connected to the mass\textendash{}radius relation of the
object \cite{key-38,key-39}. Thus, NLED effects turn out to be important
as regard to the mass-radius relation, which is maximum for a BH. 

Following this approach, in \cite{key-9} a particular non singular
exact solution of Einstein field equation has been found adapting
to the BH case the cosmological analysis in \cite{key-43}. In fact,
the conditions concerning the early era of the Universe, when very
high values of curvature, temperature and density were present \cite{key-1,key-9},
and where matter should be identified with a primordial plasma \cite{key-1,key-9},
are similar to the conditions concerning BH physics. This is exactly
the motivation because various analyses on BHs can be applied to the
Universe and vice versa \cite{key-1,key-9}. 

The model works on a homogeneous and isotropic star (a collapsing
{}``\emph{ball of dust}'') supported against self-gravity entirely
by radiation pressure. Let us consider the Heisenberg-Euler NLED Lagrangian
\cite{key-9,key-42,key-43}

\begin{equation}
\mathcal{L}_{m}\equiv-\frac{1}{4}F+c_{1}F^{2}+c_{2}G^{2},\label{eq: 3}\end{equation}

where $G=\frac{1}{2}\eta_{\alpha\beta\mu\nu}F^{\alpha\beta}F^{\mu\nu}$,
$F\equiv F_{\mu\nu}F^{\mu\nu}$ is the electromagnetic scalar and
$c_{1}$ and $c_{2}$ are constants. Through an averaging on electric
and magnetic fields \cite{key-9,key-42,key-43}, the Lagrangian (\ref{eq: 3})
enables a modified radiation-dominated equation of state ($p$ and
$\rho$ are the pressure and the density of the collapsing star) \begin{equation}
p=\frac{1}{3}\rho-\rho_{\gamma},\label{eq: star}\end{equation}

where\emph{ a quintessential density term} $\rho_{\gamma}=\frac{16}{3}c_{1}B^{4}$
is present together with the standard term $\frac{1}{3}\rho$ \cite{key-9,key-42,key-43}.
$B$ is the strength of the magnetic field associated to $F$. The
interior of the star is represented by the well-known Robertson\textendash{}Walker
line-element \cite{key-1,key-9}

\begin{equation}
ds^{2}=-dt^{2}+a(t)[d\chi^{2}+\sin^{2}\chi(d\theta^{2}+\sin^{2}\theta d\varphi^{2})].\label{eq: metrica conformemente piatta}\end{equation}

Using $\sin\chi$ we choose the case of positive curvature, which
is the only one of interest because it corresponds to a gas sphere
whose dynamics begins at rest with a finite radius \cite{key-1,key-9}.
Considering Eq. (\ref{eq: 3}) together with the stress-energy tensor
of a relativistic perfect fluid \cite{key-1,key-9,key-42,key-43}

\begin{equation}
T=\rho u\otimes u-pg,\label{eq: stress energy 2}\end{equation}
 where $u$ is the four-vector velocity of the matter and $g$ is
the metric, the Einstein field equation gives the relation \cite{key-9,key-43}

\begin{equation}
t=\int_{a(0)}^{a(t)}dz(\frac{B_{0}^{2}}{6z^{2}}-\frac{8c_{1}B_{0}^{4}}{6z^{6}}-1)^{-\frac{1}{2}},\label{eq: soluzione}\end{equation}

being $B_{0}=a^{2}B.$ The expression (\ref{eq: soluzione}) is not
singular for values of $c_{1}>0$ in Eq. (\ref{eq: 3}) \cite{key-9,key-43}.
In fact, the presence of the quintessential density term $\rho_{\gamma}$
permits to violate the reasonable energy condition \cite{key-1} of
the singularity theorems. By using elliptic functions of the first
and second kind, one gets a parabolic trend for the scale factor near
a minimum value $a_{f}$ in the final stages of gravitational collapse
\cite{key-9}.

In concrete terms, by calling $l,m,n$ the solutions of the equation
$8c_{1}B_{0}^{4}-B_{0}^{2}x+3x^{3}=0,$ reads \cite{key-9,key-43}

\begin{equation}
\begin{array}{c}
t=[-(m-l)^{\frac{1}{2}}\beta_{1}(\arcsin\sqrt{\frac{z-l}{m-l}},\sqrt{\frac{l-m}{l-n}})\\
\\+n(m-l)^{-\frac{1}{2}}\beta_{2}(\arcsin\sqrt{\frac{z-l}{m-l}},\sqrt{\frac{l-m}{l-n}})]|_{z=a^{2}(0)}^{z=a^{2}(t)},\end{array}\label{eq: soluzione-1}\end{equation}

where \begin{equation}
\beta_{1}(x,y)\equiv\int_{0}^{\sin x}dz[(1-z^{2})^{-1}(1-y^{2}z^{2})^{-1}]\label{eq: ell1}\end{equation}
 is the elliptic function of the first kind and \begin{equation}
\beta_{2}(x,y)\equiv\int_{0}^{\sin x}dz[((1-z^{2})^{-1})^{-\frac{1}{2}}((1-y^{2}z^{2})^{-1})^{\frac{1}{2}}]\label{eq: ell2}\end{equation}
 is the elliptic function of the second kind.

Then, recalling that the Schwarzschild radial coordinate, in the case
of the BH geometry (\ref{eq: metrica conformemente piatta}), is $r=a\sin\chi_{0}$
\cite{key-1,key-9}, where $\chi_{0}$ is the radius of the surface
in the coordinates (\ref{eq: metrica conformemente piatta}), one
gets a final radius of the star, if $B_{0}$ has an high strength \cite{key-9}

\begin{equation}
r_{f}=a_{f}\sin\chi_{0}>2M\label{eq: non singolare}\end{equation}
 where $M$ is the mass of the collapsed
star and $2M$ the gravitational radius in natural units \cite{key-1,key-9}.
Thus, we find that the mass of the star generates a curved space-time
without EHs.

\section{{\large Conclusion remarks}}

Black holes should be the final result of an irreversible gravitational
collapse of very massive bodies. An unsolved problem, which was present
starting by the first historical papers concerning black holes, is
the presence of a space-time singularity in their core. It is a common
opinion that this problem could be solved when a correct quantum gravity
theory will be, finally, constructed. 

In this paper we reviewed a way to remove black hole singularities
at a classical level i.e. without arguments of quantum gravity. By
using a particular non-linear electrodynamics Lagrangian, an exact
solution of Einstein field equations has been shown. The solution
prevents the collapsing object to reach the gravitational radius,
thus the final result becomes an extreme electromagnetic compact 
object exhibiting an utterly extreme gravitational redshift $z 
\longrightarrow \infty$, i.e., a black star, that is nothing else than 
an astrophysical object where both singularities and event horizons were 
removed. Such solution is not a mathematical artifice. In fact, this kind
of Lagrangian has been recently used in various analysis in astrophysics,
like surface of neutron stars and pulsars. The authors also recently
adapted the analysis on a cosmological context by showing that the
big-bang singularity can be removed too \cite{key-42}. 

Potential removal of BH horizons and singularities is an exciting
and rapidly advancing field of research on theoretical, observational
and experimental fronts. We take the chance to signal some recent
results \cite{key-44,key-45}.

\subsubsection*{Acknowledgements }

The R. M. Santilli Foundation has to be thanked for partially supporting
this paper (Research Grant of the R. M. Santilli Foundation Number
RMS-TH-5735A2310).


\begin{thebibliography}{45}
\bibitem{key-1}\foreignlanguage{italian}{C. W. Misner, K. S. Thorne
and J. A. Wheeler\textit{, Gravitation}, W. H. Feeman and Company
(1973).}

\selectlanguage{italian}%
\bibitem{key-2}P. C. W. Davies, Rep. Prog. Phys. 41, 1313\textendash{}1355
(1978).

\bibitem{key-3}K. Schwarzschild, Sitzungsber. Preuss. Akad. Wiss.
Berlin (Math. Phys.) 1916, 189-196 (1916).

\bibitem{key-4}G. D. Birkhoff\textit{, Relativity and Modern Physics}.
Cambrigdge, MA: Harvard University Press. LCCN 23008297 (1923).

\bibitem{key-5}S. Chandrasekhar, \textit{The Highly Collapsed Configurations
of a Stellar Mass}, Mont. Not. Roy. Astr. Soc. 91, 456\textendash{}466
(1931).

\selectlanguage{english}%
\bibitem[6]{key-6}\foreignlanguage{italian}{H. Nicolai, G. F. R.
Ellis, A. Ashtekar and others, Gen. Rel. Grav. 41, 4, 673-1011, \emph{Special
Issue on quantum gravity} (April 2009).}

\bibitem[7]{key-7}S. Robertson and D. Leiter, Mon. Roy. Astron. Soc.
50, 1391 (2004).

\bibitem[8]{key-8}R. Schild, D. Leiter and S. Robertson, Astron.
J. 2, 420-432 (2006).

\bibitem[9]{key-9}C. Corda and H. J. Mosquera Cuesta, Mod. Phys.
Lett A 25, 28, 2423-2429 (2010).

\bibitem[10]{key-10}A. Mitra, Phys. Rev. Lett. 81, 4774 (1998).

\bibitem[11]{key-11}A. Mitra, Phys. Rev. D, 74, 2, 024010 (2006).

\bibitem[12]{key-12}A. Mitra, Mon. Not. Roy. Astron. Soc. 369, 492
(2006).

\bibitem[13]{key-13}A. Mitra, Journ. Math. Phys. 50, 4, 042502-042503
(2009).

\bibitem[14]{key-14}J. P. S. Lemos, O.B. Zaslavskii, arXiv:1004.4651
(2010).\foreignlanguage{italian}{ }

\bibitem[15]{key-15}A. G. Agnese and M. La Camera, Phys. Rev. D 31,
1280\textendash{}1286 (1985).

\bibitem[16]{key-16}S. Robertson and D. Leiter, Astrophys. J. 596,
L203-L206 (2003).

\bibitem[17]{key-17}E. Einstein, Ann. Math. 40, 4, 922\textendash{}936
(1939).

\bibitem[18]{key-18}A. Einstein, \textit{{}``Zur allgemeinen Relativit\"{a}tstheorie'',
}\textit{\emph{Sitzungsberichte der a Koniglich Preußischen Akademie
der Wissenschaften 1915, 778-86 (1915).}}\foreignlanguage{italian}{ }

\selectlanguage{italian}%
\bibitem[19]{key-19}\foreignlanguage{english}{J. Droste, Proc. K.
Ned. Akad. Wet. 19, 197 (1917). }

\bibitem[20]{key-20}\foreignlanguage{english}{H. Weyl, Ann. Phys.
(Leipzig) 54, 117 (1917).}

\selectlanguage{english}%
\bibitem[21]{key-21}\foreignlanguage{italian}{D. Hilbert, Nachr.
Ges. Wiss. Gottingen, Math. Phys. Kl., 53 (1917).}

\selectlanguage{italian}%
\bibitem[22]{key-22}L. S. Abrams, Phys. Rev. D20, 2474 (1979), also
in arXiv:gr-qc/0201044.

\bibitem[23]{key-23}\foreignlanguage{english}{S. Antoci and D.E.
Liebscher, Gen. Rel. Grav. 35, 5, 945-950 (2003).}

\bibitem[24]{key-24}L. S. Abrams, Can. J. Phys. 67, 919 (1989), also
in arXiv:gr-qc/0102055. 

\bibitem[25]{key-25}S. Antoci, in \char`\"{}Meteorological and Geophysical
Fluid Dynamics (a book to commemorate the centenary of the birth of
Hans Ertel)\char`\"{}, W. Schroder Editor, Science Edition, Bremen,
2004 , also in arXiv:physics/0310104.\foreignlanguage{english}{ }

\bibitem[26]{key-26}C. Corda, arXiv:1010.6031, to be published in
Electronic Journal of Theoretical Physics 

\bibitem[27]{key-27}A. S. Eddington, Nature 113, 192 (1924). 

\selectlanguage{english}%
\bibitem[28]{key-28}A. Einstein and G. Lemaître\foreignlanguage{italian}{,
TALKING OVER THE UNIVERSE AND ITS ORIGIN, Lectures in Pasadena, California,
JWC-CX 1-13-33 {[}January 13, 1933{]} WC MET. (STAMPED JAN 14 1933).}

\bibitem[29]{key-29}\foreignlanguage{italian}{J. R. Oppenheimer and
G. M. Volkoff, Phys. Rev. 55, 4, 374\textendash{}381 (1939).}

\bibitem[30]{key-30}\foreignlanguage{italian}{R. Ruffini and J. A.
Wheeler, Phys. Tod. 30\textendash{}41 (1971).}

\bibitem[31]{key-31}D. Finkelstein, Phys. Rev. 110, 965\textendash{}967
(1958). 

\bibitem[32]{key-32}M. Kruskal, Phys. Rev. 119, 1743\textendash{}1745
(1960). 

\bibitem[33]{key-33}A. Hewish, S. J. Bell and J. D. H Pilkington,
Nature 217, 709\textendash{}713 (1968). 

\bibitem[34]{key-34}R. Kerr, Phys. Rev. Lett. 11, 237\textendash{}238
(1963). 

\bibitem[35]{key-35}E. T. Newman and A. Janis, Journ. Math. Phys.
6, 6, 915\textendash{}917 (1965). 

\bibitem[36]{key-36}S. Hawking, \emph{A Brief History of Time, }Bantam
Dell Publishing Group (1988).

\bibitem[37]{key-37}A. Ewing, Science News Letter of 18 January
(1964).

\selectlanguage{italian}%
\bibitem[38]{key-38}\foreignlanguage{english}{H. J. Mosquera Cuesta
and J. M. Salim, Ap. J. 608, 925-929 (2004). }

\bibitem[39]{key-39}\foreignlanguage{english}{H. J. Mosquera Cuesta
and J. M. Salim, Mont. Not. Roy. Ast. Soc. 354, L55-L59 (2004).}

\bibitem[40]{key-40}Private communication with \foreignlanguage{english}{Professor
Darryl Jay Leiter.}

\bibitem[41]{key-41}S. A. Wheeler\foreignlanguage{english}{ and I.
Ciufolini, \emph{Gravitation and inertia}, Princeton University Press
(1995). }

\bibitem[42]{key-42}\foreignlanguage{english}{C. Corda and H. J.
Mosquera Cuesta, \emph{Astropart. Phys.} \textbf{34}, 587 (2011).}

\selectlanguage{english}%
\bibitem[43]{key-43}V. A. De Lorenci, R. Klippert, M. Novello and
J. M. Salim, \emph{Phys. Rev. D} \textbf{65}, 063501 (2002).

\bibitem[44]{key-44}A. Mitra, Astrophys. Sp. Sci. Lett. 332, 43-48
(2011).

\bibitem[45]{key-45}S. Yi, JCAP 1101, 031 (2011).
\end{thebibliography}
\end{document}